\documentstyle[preprint,aps,prl]{revtex}

\begin{document}

\draft

\title{ Scaling and thermodynamics of a trapped Bose-condensed gas} 

\author{ S. Giorgini$^{1}$, L. P. Pitaevskii$^{2,3}$ and S. Stringari$^{1}$}

\address{$^{1}$Dipartimento di Fisica, Universit\`a di Trento, \protect\\
and Istituto Nazionale di Fisica della Materia, I-38050 Povo, Italy}
\address{$^{2}$Department of Physics, Technion, 32000 Haifa, Israel}
\address{$^{3}$Kapitza Institute for Physical Problems, 117454 Moscow, Russia}


\maketitle

\begin{abstract}

{\it We investigate the thermodynamics of  a Bose gas interacting
with repulsive forces and  confined in a 
harmonic trap. We show that 
the relevant parameters of the system (temperature, number $N$ of
atoms, harmonic oscillator length, deformation of the trap, $s$-wave
scattering length) fix its large $N$ 
thermodynamic behaviour through two dimensionless
scaling parameters. These  are 
the reduced temperature $t=T/T^0_c$
and the ratio $\eta$ between 
the $T=0$ value 
of the chemical potential, evaluated in the Thomas-Fermi limit, and
the critical temperature $T_c^0$ of the 
non-interacting model.
The scaling functions relative to the
condensate fraction, energy, chemical potential and moment of inertia 
are calculated within the Popov approximation. } 

\end{abstract}


\narrowtext

First measurements of relevant thermodynamic properties in  gases of
alkali atoms confined in magnetic traps have been recently become
available \cite{MIT,JIL}. These include the temperature dependence of the 
fraction of atoms in the condensate 
and the release energy.
On the theoretical side the study of non uniform  interacting Bose gases
has a long history, starting from the works by Gross,
Pitaevskii and Fetter \cite{TH1} and the first approaches to thermodynamics 
developed 
in ref. \cite{HF}. More recently several theoretical papers have 
focused on the problem of the thermodynamic 
behaviour  of such systems \cite{TH2,GRI,TH3,US,HZG,SZ}, and first calculations
of the temperature dependence of the condensate have become available 
\cite{TH3,US,HZG}.

The properties of these
trapped Bose gases at a given temperature are fixed by a large number
of parameters: total number of atoms in the trap, atomic mass, 
harmonic oscillator length,
deformation of the trap and strength of the interaction, which in a dilute gas
is fixed by the $s$-wave scattering length. This makes these systems
very rich for both experimental and theoretical investigations, but at the
same time difficult to study in a systematic way. For example
the deformation of the trap in the experiments carried out at MIT 
\cite{MIT} 
and at Jila \cite{JIL} as well as the corresponding number of atoms are quite
different and it is not always easy to compare the experimental results.
The purpose of this letter is to point out the occurrence of  a scaling 
behaviour in the thermodynamics of these many-body systems. This permits
to compare very different experimental situations, provided they 
correspond to the same value of the scaling parameters.

A {\it first} natural scaling parameter is given by the reduced temperature
$t=T/T_c^0$,
where $k_BT_c^0 = \hbar \omega (N/\zeta(3))^{1/3}$ \cite{IDG}, 
with $\zeta(3)\simeq 1.202$, 
is the critical temperature predicted by the noninteracting harmonic oscillator
model in the large $N$ limit and 
$\omega=(\omega_x\omega_y\omega_z)^{1/3}$ is the geometrical average of the
oscillator frequencies characterizing the confining potential 
$V_{ext} = {1\over 2}m(\omega^2_xx^2 + \omega^2_yy^2 + \omega^2_zz^2)$.

In addition to the reduced temperature $t$ we 
introduce a {\it second} scaling parameter accounting for the role 
of the two-body repulsive interaction. This parameter is fixed by 
the ratio $\eta$ between the $T=0$  
value of the chemical potential, calculated in the Thomas-Fermi approximation,
and the critical temperature $k_BT_c^0$. In the Thomas-Fermi
approximation
the ground state density of a Bose gas takes the form \cite{TF} 
$n_0({\bf r})=(\mu_0^{TF}-V_{ext}({\bf r}))/g$,
where $g=4\pi\hbar^2a/m$
is the coupling constant fixed by the (positive) 
$s$-wave scattering length $a$ and the 
chemical potential is given by $\mu_0^{TF} = 
{1\over2}\hbar\omega(15 N a/a_{HO})^{2/5}$,
where
$a_{HO}=(\hbar/m\omega)^{1/2}$ is the harmonic oscillator length.
The parameter $\eta$ then takes the useful form
\begin{eqnarray}
\eta = {\mu_0^{TF} \over k_BT_c^0} = \alpha \left( N^{1/6}\frac{a}{a_{HO}}
\right)^{2/5} \;\;, 
\label{eta}
\end{eqnarray}
with $\alpha=\zeta(3)^{1/3}15^{2/5}/2\simeq1.57$.
Note that $\eta$ exhibits a very smooth dependence on $N$
($\eta \sim N^{1/15}$) 
and its value in the presently available experiments ranges 
from $0.38$ to $0.45$ even if $N$ varies by orders
of magnitute in the different experiments.
The ratio (\ref{eta}) can be also written in the form  
$\eta=2.24 (a^3n_0({\bf r}=0))^{1/6}$,  
where
$n_0({\bf r}=0)$ is the density of the atomic cloud in the center of the trap 
evaluated at zero temperature. 
With the above values of $\eta$,  
the gas parameter $a^3n_0({\bf r}=0)$ turns out to be extremely small 
(10$^{-4}$-10$^{-5}$) so that the system is always very dilute.

In the following we will show that the relevant thermodynamic properties 
of the system exhibit, when expressed in their natural units, a scaling 
behaviour in terms of the parameters $t$ and $\eta$. 
Scaling is actually achieved only if $N$ is sufficiently
large and finite size corrections to the asymptotic behaviour 
of the various thermodynamic functions can be ignored.
We will show that scaling is reached with very good accuracy for most
thermodynamic properties not only in the experimental conditions of ref.
\cite{MIT}  
where $N$ is of order of $10^6-10^7$, but also in the ones
of ref. \cite{JIL}  where $N$ is much smaller ($10^4-10^5$).

The scaling behaviour can be  exploited 
by investigating the thermodynamic behaviour of these systems
within the Popov approximation \cite{POP,GRI,US}.
This approximation provides coupled equations for the condensate density
and the thermal density of the system. In this letter we will use 
the semiclassical approximation for the excited states and hence we
will always consider temperatures much larger than the oscillator temperature
$\hbar\omega/k_B$. The equation for the condensate
is given by
\begin{equation}
- \frac{\hbar^2}{2m}\nabla^2\Phi({\bf r}) + [ V_{ext}({\bf r}) - \mu  
+ g(n_0({\bf r}) + 2n_T({\bf r})) ]\Phi({\bf r}) = 0 ,
\label{cequ}
\end{equation}
and the dispersion law of the elementary excitations takes the 
Bogoliubov-type form \cite{US} 
\begin{equation}
\epsilon({\bf p},{\bf r}) = \sqrt{\left( \frac{p^2}{2m} + V_{ext}({\bf r})
- \mu + 2gn({\bf r}) \right)^2 - g^2n_0^2({\bf r})} . 
\label{elex}
\end{equation}
In eqs. (\ref{cequ}), (\ref{elex}) $n_0=\Phi^2$ is the density 
of the condensate, 
$n = n_T +n_0$ is the total density of the system and the
thermal density $n_T$ is defined by 
$n_T({\bf r}) = 
\int d{\bf p}/(2\pi\hbar)^3 (-\partial\epsilon/\partial\mu) f(\beta\epsilon)$
where  
$f(\beta\epsilon) = (exp(\beta\epsilon)-1)^{-1}$ is the quasi-particle 
distribution function. 
The semiclassical approximation (\ref{elex}) to the excitation 
spectrum is accurate  
only if the non-condensate density $n_T$ varies smoothly on the scale
of the oscillator length $a_{HO}$, and if the relevant values of $p$
satisfy the condition $p \gg \hbar/a_{HO}$, which also implies $\epsilon \gg 
\hbar\omega$.
The above approximation, called Popov approximation,
coincides with the Bogoliubov approach at very low temperatures,
while at high $T$ it coincicides with the finite temperature Hartree-Fock
theory for Bose systems \cite{HF} (for further details on the Popov 
approximation see refs. \cite{GRI,US}). 
Equations (\ref{cequ})-(\ref{elex}) are solved \cite{US} using a 
self consistent procedure, the value
of the chemical potential $\mu$ being determined by the normalization
condition 
$\int d{\bf r}(n_0({\bf r})+n_T({\bf r})) = N_0+N_T=N$.

The scaling behaviour of the Popov equation is exhibited in the 
Thomas-Fermi limit, which is reached when the ratio $\mu_0^{TF}/\hbar\omega =
0.94 N^{1/3}\eta$ is much larger than unity. In this limit, which is formally
achieved taking $N\rightarrow\infty$ with $\eta$ fixed, the kinetic energy 
term in the Schr\"odinger equation (\ref{cequ}) for the condensate can be 
neglected. 
By introducing the dimensionless quantities 
$\tilde{p}_i = \sqrt{1/2mk_BT_c^0}p_i$, $\tilde{r}_i=\sqrt{m/2k_BT_c^0}
\omega_i r_i$,
$\tilde{\mu} = \mu/k_BT_c^0$, $\tilde{\epsilon} = \epsilon/k_BT_c^0$, 
$\tilde{n}_T=n_T(2k_BT_c^0/m\omega^2)^{3/2}/N$ and 
$\tilde{n}_0=n_0(2k_BT_c^0/m\omega^2)^{3/2}/N$,
one can rewrite the Popov equations in the reduced form
\begin{equation}
\tilde{n}_0(\tilde{r}) = \frac{1}{\tilde{g}} \left( \tilde{\mu} 
- \tilde{r}^2 - 
2\tilde{g}\tilde{n}_T \right) 
\;\theta(\tilde{\mu}-\tilde{r}^2-2\tilde{g}\tilde{n}_T) \;\;,
\label{red1}
\end{equation}
\begin{equation}
\tilde{\epsilon}(\tilde{p},\tilde{r}) = \sqrt{\left(\tilde{p}^2
+\tilde{r}^2-\tilde{\mu}
+2\tilde{g}(\tilde{n}_0+\tilde{n}_T)\right)^2
-\tilde{g}^2\tilde{n}_0^2} \;\;, 
\label{red2}
\end{equation}
where 
$\tilde{n}_T(\tilde{r}) =  
1/(\pi^3\zeta(3))\int d\tilde{{\bf p}} 
(-\partial\tilde{\epsilon}/\partial\tilde{\mu})
f(\tilde{\epsilon}/t)$, 
$\tilde{g}=8\pi\eta^{5/2}/15=5.2 N^{1/6}a/a_{HO}$ and   
$\theta(x)$ is the step function. 
The normalization condition for the reduced densities reads:
$\int d\tilde{\bf r} (\tilde{n}_0+\tilde{n}_T) = 1$.
Equations (\ref{red1})-(\ref{red2}) exhibit the anticipated 
scaling behaviour in the variables $t$ and $\eta$. Starting from their 
solutions
one can calculate 
the condensate and the thermal densities as well as the 
excitation spectrum. This gives access to 
all the relevant thermodynamic quantities of the system.
For example the condensate fraction is given by
\begin{equation}
{N_0 \over N} = 1 - 
\frac{1}{\pi^3\zeta(3)}\int\; d\tilde{{\bf r}} d\tilde{{\bf p}} 
\left(-\frac{\partial\tilde{\epsilon}}{\partial\tilde{\mu}}\right)
f(\tilde{\epsilon}/t) \;\;.
\label{N0red}
\end{equation}
In particular the vanishing of the right hand side of eq. (\ref{N0red}) fixes 
the value of the
critical temperature $t_c=T_c/T_c^0$.
To the lowest order in the renormalized 
coupling constant $\tilde{g}$ one finds \cite{US,SZ}
$t_c \simeq 1 - 0.43 \eta^{5/2} = 1-1.3 N^{1/6}a/a_{HO}$.
The energy of the system can be calculated starting
from the standard thermodynamic relation in terms of the entropy.
In units of $Nk_BT_c^0$ one finds:
\begin{equation}
{E\over Nk_BT_c^0} = \frac{5}{7}\eta + \int_0^t\;dt' t'\frac{\partial s}
{\partial t'} \;\;, 
\label{Ered}
\end{equation}
where the entropy per particle $s$ is given by the combinatorial expression
\begin{eqnarray}
s(t,\eta) &=& \frac{k_B}{\pi^3\zeta(3)} \int\;
d\tilde{{\bf r}}d\tilde{{\bf p}} [ (1+f(\tilde{\epsilon}/t))
\log(1+f(\tilde{\epsilon}/t)) \nonumber \\
&-& f(\tilde{\epsilon}/t)\log 
f(\tilde{\epsilon}/t) ]
\;\;.
\label{sred}
\end{eqnarray}
Another important quantity is 
the normal (non superfluid) density given by the natural generalization
\begin{equation}
\rho_n({\bf r}) = - \int\;\frac{d{\bf p}}{(2\pi\hbar)^3} 
\frac{p^2}{3} \frac{\partial f(\beta\epsilon)}{\partial\epsilon} 
\label{rhon}
\end{equation}
of the Landau formula \cite{LL} 
to the case of non uniform 
systems. 
Note that in general the normal density $\rho_n$ differs from the
non condensate density $n_T$. Only at  temperatures 
larger than the chemical potential one has $\rho_n \simeq m n_T$.
In terms of $\rho_n$
one can calculate the moment of inertia, defined as  the linear response
function to a crancking rotational field \cite{STR}. In the case of an axially
symmetric trap one finds the result
\begin{equation}
\Theta = \int\;d{\bf r} (x^2+y^2)\rho_n({\bf r}) \;\;.
\label{Theta}
\end{equation}
Deviations of $\Theta$ from the rigid value $\Theta_{rigid} = m \int d{\bf r}
(x^2+y^2) n({\bf r})$
provide a signature of the superfluid behaviour of the system.
Above $T_c$ one has $\rho_n=mn_T=mn$ and $\Theta=\Theta_{rigid}$.

In Fig.1 we explicitly show the accuracy of the scaling behaviour, by 
plotting 
the scaling function (\ref{N0red}) calculated with $\eta=0.45$ and  
the results obtained from the full solution
of the Popov equations (\ref{cequ})-(\ref{elex}), 
with different choices of $N$ and of the other
parameters, yielding the same value $\eta$. 
The open circles have been obtained with the choice: 
$a/a_{HO}=7.35\times10^{-3}$, 
$\lambda=\omega_z/\omega_x=\sqrt{8}$, $N=5\times 10^4$, which corresponds to 
the
experimental conditions of ref. \cite{JIL}, while
the solid circles have been calculated with the choice: $a/a_{HO}=2.55\times
10^{-3}$, $\lambda=18/320$, $N=2.9\times10^7$, which is close to the 
experimental situation of ref. \cite{MIT}. 
This example shows how extremely different experimental conditions give rise 
to the same thermodynamic behaviour.
The figure clearly
shows that scaling is very well verified for these configurations. 
Only very 
close to $T_c$ the $N=5\times 10^4$ points exhibit small deviations from 
the scaling behaviour. In fact close to $T_c$ the scaling law (\ref{N0red}) 
is approached more slowly with increasing $N$.
In the same figure we also plot the predictions for the same configurations
obtained switching off the two-body interaction. Deviations from the 
non-interacting result $1-t^3$ are due to finite size effects, whose
importance, in the presence of the interaction, turns out to be strongly
quenched.
The scaling behaviour has been verified also for the other 
values of the parameter $\eta$ and for all
the other thermodynamic properties considered in this work.

In Fig. 2 we present results for the condensate fraction $N_0/N$
as a function of the reduced temperature $t$ for three different values
of the scaling parameter $\eta$, 
covering the presently available 
experimental conditions. The open diamonds with the error bars are the 
results of the Monte-Carlo simulation of ref. \cite{TH3} which correspond to 
the value $\eta=0.33$ and which are in good agreement with our 
predictions. The dots are the experimental results of ref. \cite{JIL}. In the
experiments the number of particles $N$ varies with $T$, with the value of 
$\eta$ ranging from 0.45 to 0.39. The experiments exhibit smaller deviations 
from the non-interacting curve with respect to the theoretical predictions.
One should however keep in mind that the measured value of $T$ corresponds to
the temperature of the thermal cloud after expansion. The identification of 
this 
temperature with the one of the system before expansion ignores 
the interaction with the condensate which is expected to produce 
an acceleration of the thermal cloud.
A preliminary estimation shows that for $T\simeq 0.5 T_c^0$ the 
final kinetic energy of the 
thermal cloud is about 10\% larger than its value before the expansion. 

In Fig.3 we present the results for the chemical potential 
in units of $k_BT_c^0$ corresponding to 
the same values of $\eta$. Notice that for $t\rightarrow 0$ the plotted 
quantity coincides with $\eta$ (see eq. (\ref{eta})).
In the classical limit, $T \gg T_c^0$, the dependence on the interaction 
parameter
$\eta$ disappears and one finds the classical ideal gas prediction
$\mu/k_BT_c^0=t\log(\zeta(3)/t^3)$.

In Fig.4 we report the results for the total energy $E$. At high temperature
the behaviour is given by the classical law $E/k_BT_c^0 = 3t$.
In Fig. 4 (inset) we also show our predictions for the release energy $E_R$,
defined as the energy of the system after the trap potential has been 
switched off. In the classical non-interacting limit one has $E_R=3Nk_BT/2$.
First experimental results for the release energy have been obtained in 
ref. \cite{JIL} and are shown in the figure.  

Finally in Fig.5 we show our predictions for the moment of inertia. We 
find that, differently from the other quantities discussed above, the ratio
$\Theta/\Theta_{rig}$ does not exhibit  any significant dependence on $\eta$. 
In the same figure we also show the moment of inertia calculated in the
non interacting model for two different values of $N$ and deformation
$\lambda$. It is worth noting that in the absence 
of interaction the value of $\Theta/\Theta_{rig}$ depends 
rather crucially on $N$ and $\lambda$.

In conclusion we have explored the scaling behaviour
of the thermodynamic properties of a trapped Bose gas. The new key
parameter which permits to discuss the role of two-body interactions is 
the scaling parameter 
(\ref{eta}), given by the value of the zero-temperature chemical potential
in units of $k_BT_c^0$. Scaling is predicted to work very
well in the typical conditions of available experiments. 

We would like to thank D. S. Jin for providing the experimental results on 
the condensate fraction and release energy \cite{JIL}. One of us (S.G.) would
like to thank the hospitality of JILA while part of this work was performed.

\begin{figure}
\caption{Condensate fraction as a function of $T/T_c^0$ for 
$\eta=0.45$ (solid line). The open circles 
refer
to $N=5\times 10^4$ Rb atoms in the JILA-type trap. The solid circles 
correspond
to $N=2.9\times 10^7$ Na atoms in the MIT-type trap.
The dotted line is the $1-t^3$ curve of the non-interacting model in 
the large $N$ limit. The open and solid triangles correspond   
to $N=5\times 10^4$ and $N=2.9\times 10^7$ non-interacting particles 
in the JILA and MIT-type traps respectively.}
\end{figure}

\begin{figure}
\caption{Condensate fraction as a function of $T/T_c^0$ for 
three values of the scaling parameter $\eta$. Solid line: 
$\eta=0.45$, long-dashed line: $\eta=0.39$, short-dashed line: $\eta=0.31$.
Open diamonds: PIMC results of ref. \protect\cite{TH3}. 
Solid circles: experimental 
results from ref. \protect\cite{JIL}.  
The dotted line refers to the non-interacting model in the large $N$  
limit.}
\end{figure}

\begin{figure}
\caption{Chemical potential as a function of $T/T_c^0$ for
three values of the scaling parameter $\eta$ (see Fig. 2). 
The dotted line refers to the non-interacting model in the large $N$ 
limit.}
\end{figure}

\begin{figure}
\caption{Total energy of the system as a function of 
$T/T_c^0$ for three values of the scaling parameter $\eta$ (see Fig. 2). 
(inset) Release energy for the same values of $\eta$. 
The solid circles are the experimental results of ref. \protect\cite{JIL}.} 
The dotted lines refer to the non-interacting model in the large $N$ limit.
\end{figure}

\begin{figure}
\caption{Ratio $\Theta/\Theta_{rigid}$ as a function of  
$T/T_c^0$ for three values of the scaling parameter $\eta$ (see Fig. 2). 
The three curves coincide almost  
exactly and are represented by the solid 
line. The dotted line refers to $N=5\times 10^4$ atoms in the JILA-type trap 
in the non-interacting model. The dot-dashed 
line refers to $N=2.9\times 10^7$ atoms in the MIT-type trap  
in the 
non-interacting model.}
\end{figure}


\begin{references}

\bibitem{MIT} M.-O. Mewes, M.R. Andrews, N.J. van Druten, D.M. Kurn, D.S. 
Durfee and W. Ketterle, Phys. Rev. Lett. {\bf 77}, 416 (1996).

\bibitem{JIL}  
J.R. Ensher, D.S. Jin, M.R. 
Matthews, C.E. Wieman and E.A. Cornell, preprint.  

\bibitem{TH1} L.P. Pitaevskii, Zh. Eksp. Teor. Fiz. 
{\bf 40}, 646 (1961) 
[Sov. Phys. JETP {\bf 13}, 451 (1961)]; E.P. Gross, Nuovo Cimento {\bf 20}, 
454 (1961); 
A.L. Fetter, Ann. Phys. 
(N.Y.) {\bf 70}, 67 (1972).

\bibitem{HF} 
V.V. Goldman, I.F. Silvera and A.J. Leggett, Phys. Rev. B 
{\bf 24}, 2870 (1981); D.A. Huse and E.D. Siggia, J. Low Temp. Phys. {\bf 46},
137 (1982).

\bibitem{TH2} J. Oliva, Phys. Rev. B {\bf 39}, 4197 (1989); 
T.T. Chou, C.N. 
Yang and L.H. Yu, Phys. Rev. A {\bf 53}, 4257 (1996);
cond-mat/9605058.

\bibitem{GRI} A. Griffin, Phys. Rev. B {\bf 53}, 9341 (1996);

\bibitem{TH3} 
W. Krauth, Phys. Rev. Lett. {\bf 77}, 3695 (1996).

\bibitem{US} S. Giorgini, L.P. Pitaevskii and S. Stringari, Phys. Rev. A 
{\bf 54}, 4633 (1996). 

\bibitem{HZG} D.A.W. Hutchinson, E. Zaremba and A. Griffin, cond-mat/9611023.

\bibitem{SZ} H. Shi and W.-M. Zheng, cond-mat/9609241.

\bibitem{IDG} S.R. de Groot, G.J. Hooyman and C.A. Ten Seldam, Proc. R. Soc. 
(London), {\bf A 203}, 266 (1950); V. Bagnato, D.E. Pritchard and D. Kleppner,
Phys. Rev. A {\bf 35}, 4354 (1987).
 
\bibitem{TF} M. Edwards and K. Burnett, Phys. Rev. A {\bf 51}, 1382 (1995); 
G. Baym and C. Pethick, Phys. Rev. Lett. {\bf 76}, 6 (1996);
F. Dalfovo and S. Stringari, Phys. Rev. A {\bf 53}, 2477 (1996); E. Timmermans,
P. Tommasini and K. Huang, cond-mat/9609234.

\bibitem{POP} V.N. Popov, {\it Functional integrals and collective excitations}
(Cambridge University Press, Cambridge, 1987).

\bibitem{IG1} S. Grossman and M. Holthaus, Phys. Lett. {\bf A 208},188 (1995) 
and Zeit. f. Naturforsch. {\bf 50 a}, 323 (1995); W. Ketterle and N.J. van 
Druten, Phys. Rev. A {\bf 54}, 656 (1996); K. Kirsten and D.J. Toms, 
Phys. Rev. A {\bf 54}, 4188 (1996); H. Haugenrud, T. Haugset and F. Ravndal, 
cond-mat/9605100.

\bibitem{LL} See for example E.M. Lifshitz and L.P. Pitaevskii, {\it 
Statistical Physics} (Pergamon, Oxford, 1980), Part 2.

\bibitem{STR} S. Stringari, Phys. Rev. Lett. {\bf 76}, 1405 (1996).
 
\end{references}
\end{document}